\documentclass[a4paper,11pt]{article}
\usepackage[T1]{fontenc}
\usepackage[utf8]{inputenc}
\usepackage{fancyhdr}
\usepackage[english]{babel}
\usepackage{booktabs}
\usepackage{amsmath}
\usepackage{slashed}
\usepackage{bbold}
\usepackage{geometry}
\usepackage{subfig}
\usepackage[titletoc]{appendix}
\usepackage{graphicx}
\usepackage{tikz}
\usepackage{dsfont}
\usetikzlibrary{chains, shapes.misc}
\usetikzlibrary{matrix,shapes,arrows,positioning,chains}
\usetikzlibrary{calc}
\usepackage[autostyle]{csquotes}

\begin{document}
\sffamily
 
\begin{center}
{\LARGE
Developing and testing the density of states FFA method\\ 
\vskip2mm
in the SU(3) spin model}
\vskip15mm
Mario Giuliani, Christof Gattringer, Pascal T\"orek
\vskip8mm
Universit\"at Graz, Institut f\"ur Physik, Universit\"atsplatz 5, 8010 Graz, Austria
\end{center}
\vskip8mm

\begin{abstract}
The Density of States Functional Fit Approach (DoS FFA) is a recently proposed modern 
density of states technique suitable for calculations in lattice field theories with a complex action problem. 
In this article we present an exploratory implementation of DoS FFA for the SU(3) spin system at 
finite chemical potential $\mu$ -- 
an effective theory for the Polyakov loop. This model has a complex action problem similar to the one 
of QCD but also allows for a dual simulation in terms of worldlines where the complex action problem is solved.
Thus we can compare the DoS FFA results to the reference data from the dual simulation and assess the 
performance of the new approach. We find that the method reproduces the observables from the dual simulation
for a large range of $\mu$ values, including also phase transitions,
illustrating that DoS FFA is an interesting approach for exploring phase diagrams
of lattice field theories with a complex action problem. 
\end{abstract}

\vskip8mm

\section{Introduction}

An ab initio lattice simulation of QCD at finite density is one of the great current challenges in lattice field theory. 
The reason is that for many field theories the action $S$ becomes complex at finite chemical potential
$\mu$ and consequently the Boltzmann factor $e^{-S}$ does not allow for a probabilistic interpretation which
is necessary for a Monte Carlo Simulation. Over the years various methods to overcome this so-called ''complex 
action problem'' have been developed. Examples are complex Langevin, the Lefshetz thimble,
Taylor expansion around $\mu=0$, 
analytical continuation to imaginary chemical potential or the dual approach, where the system is rewritten 
to new variables such that all weights are real and positive. In the end, probably only the combination of the results from 
different approaches will lead to a reliable understanding of finite density lattice QCD.

Another approach are Density of States (DoS) methods which appear in the literature early on  
\cite{Gocksch:1987nt,Gocksch:1988iz} and were revisited regularly, see, e.g., 
\cite{Schmidt:2005ap,Fodor:2007vv,Ejiri:2007ga,Ejiri:2012ng}. For the application of DoS techniques 
to finite density lattice field theory the main challenge is to compute the density of states $\rho(x)$ with very high 
precision, since for the evaluation of observables it is integrated over with a highly fluctuating function. 

An important new technique for pushing up the precision for the density, related to a proposal by 
Wang and Landau \cite{PhysRevLett.86.2050}, was introduced by Langfeld, Lucini and Rago in 
\cite{Langfeld:2012ah,Langfeld:2013xbf,Langfeld:2014nta,Langfeld:2015fua,Garron:2016noc}. The key idea of this 
so-called LLR method is to divide the variable 
$x$ of the density $\rho(x)$ into small intervals and to determine the variation of $\rho(x)$ in each interval using restricted 
vacuum expectation values. These new techniques make it plausible that DoS methods could become competitive
also for the analysis of finite density lattice QCD \cite{Gattringer:2016kco}. 

Here we present an exploratory study of a variant of DoS techniques, the Density of States Functional Fit Approach
(DoS FFA) \cite{Mercado:2014dva,Gattringer:2015lra,Gattringer:2015eey}. 
Similar to the LLR approach we parameterize the density on small intervals and evaluate restricted vacuum 
expectation values. These depend on a free parameter $\lambda$, and when using the density $\rho(x)$ one
can derive a closed form for their functional dependence on $\lambda$. With a one-parameter fit of the Monte Carlo data
one then can precisely determine $\rho(x)$ on the corresponding interval. The Dos FFA has been introduced and tested 
for the $\mathds{Z}_3$ spin system and was found to reproduce the results of a reference simulation 
with dual variables for a relatively wide range of parameters \cite{Mercado:2014dva,Gattringer:2015lra,Gattringer:2015eey}.

In this article we develop the DoS FFA further and
apply it to a system with continuous degrees of freedom, the SU(3) spin model with chemical potential 
(the $\mathds{Z}_3$ model has discrete degrees of freedom). The SU(3) spin model is an interesting candidate theory 
since it is an effective model for the Polyakov loop and inherits the complex action problem from QCD. 
Maybe even more important 
is the fact that the model has a dual representation in terms of worldlines \cite{Gattringer:2011gq}, such that 
only real and positive weights appear. Thus one can run Monte Carlo simulations directly in terms of the
dual variables \cite{Mercado:2012ue,Delgado:2012uh} and in this way generate reference data for assessing the
efficiency and application range of DoS FFA. 

In our exploratory study we present the generalization of the DoS FFA to continuous variables, discuss the
implementation for the SU(3) spin model and use the reference data from the dual simulation for evaluating the 
new DoS approach. In addition we discuss various techniques that allow for an optimal use of the numerical resources 
in DoS calculations.

\section{Definition of the model and the density of states}

Using lattice QCD as a starting point one can apply strong coupling expansion for the Wilson gauge action 
and a hopping expansion of the fermion determinant of the Wilson-Dirac operator at finite chemical potential to 
obtain an effective action for the Polyakov loop. The resulting model is the so-called SU(3) spin model with action
\begin{equation}
S[\mathrm{P}] \; = \; - \, \tau \, \sum\limits_{\vec{n} \in \Lambda} \sum\limits_{\nu=1}^3 
\Bigl[ \mathrm{P}(\vec{n}) \, \mathrm{P}(\vec{n}+\vec{\nu})^{\dagger} + h.c. \Bigr] \; - \;  
\kappa \sum\limits_{\vec{n} \in \Lambda } 
\Bigl[ e^\mu \, \mathrm{P}(\vec{n}) + e^{-\mu} \, \mathrm{P}(\vec{n})^{\dagger} \Bigr]  \; ,
\label{action}
\end{equation}
where the dynamical degrees of freedom $\mathrm{P}(\vec{n})$ are the traces of SU(3) matrices representing
the Polyakov loop, which live on the sites $\vec{n}$ of a 3-dimensional lattice $\Lambda$ with periodic 
boundary conditions. The nearest neighbor coupling $\tau$ is an increasing function of the temperature 
$T$ in the underlying lattice QCD theory. $\kappa$ is a decreasing function of the quark mass and is 
proportional to the number of (mass degenerate) flavors. $\mu$ is the chemical potential, and we have absorbed the 
inverse temperature of the underlying lattice QCD such that our $\mu$ corresponds to the dimensionless combination 
$\mu/T$ of QCD. 

The partition function is obtained by integrating the Boltzmann factor $e^{-S[\mathrm{P}]}$ over all configurations
of the dynamical degrees of freedom $\mathrm{P}(\vec{n})$ where the path integral measure is the product over the 
Haar measures on all sites $\vec{n}$,
\begin{equation}
Z  \; = \; \int \!\! \mathcal{D}[\mathrm{P}] \; e^{-S[\mathrm{P}]} \qquad 
\text{where} \qquad \mathcal{D}[\mathrm{P}] \; = \; \prod_{\vec{n}} d \, \mathrm{P}(\vec{n}) \;.
\end{equation}
Since the action depends only on the trace of the SU(3) matrices we need only  
two angles $\theta_1(\vec{n}), \theta_2(\vec{n}) \in [-\pi,\pi]$ to parameterize our degrees of freedom,
\begin{equation}
\mathrm{P}(\vec{n}) \; = \; \mathrm{Tr} \, \mathrm{diag} \; \Big( 
e^{i \theta_1(\vec{n})} , e^{i \theta_2(\vec{n})} ,  e^{-i(\theta_1(\vec{n})+\theta_2(\vec{n}))} \Big) \; = \; 
e^{i \theta_1(\vec{n})} + e^{i \theta_2(\vec{n})} + e^{-i(\theta_1(\vec{n})+\theta_2(\vec{n}))} \; ,
\end{equation}
and we can work with the reduced Haar measure 
\begin{equation}
d \, \mathrm{P}(\vec{n}) = \sin\! \left( \! \frac{\theta_1(\vec{n}) - \theta_2(\vec{n})}{2} \! \right)^{\!2} 
\sin\! \left( \! \frac{2 \theta_1(\vec{n}) + \theta_2(\vec{n})}{2} \! \right)^{\!2} 
\sin\! \left( \! \frac{\theta_1(\vec{n}) + 2 \theta_2(\vec{n})}{2} \! \right)^{\!2} d \theta_1(\vec{n}) \, d \theta_2(\vec{n}) \, .
\end{equation}

It is obvious that for finite chemical potential $\mu \neq 0$, the action (\ref{action}) 
has a non-zero imaginary part such that the Boltzmann factor $e^{-S[\mathrm{P}]}$ has a phase and the model 
in the conventional representation has a complex action problem. However, as already mentioned in the discussion, 
the SU(3) spin model has a dual representation in terms of worldlines \cite{Gattringer:2011gq} where the complex
action problem is solved and Monte Carlo simulations become possible for arbitrary 
$\mu$ \cite{Mercado:2012ue,Delgado:2012uh}. These
dual results will be used as reference data for the assessment of the DoS FFA.
\vskip5mm

The first step of a density of states method is to define the density to be used. Here we work with 
a weighted density and for 
its definition the action is divided into a real and an imaginary part, i.e., $S[P]=S_R[P]+iS_I[P]$. 
Only the second part of (\ref{action}) 
contains $\mu$ and we can write,
\begin{equation}
\label{eq:Action-imaginary}
\hspace*{-0.35cm}
\kappa \, \sum_{\vec{n}} \bigl[ e^{\mu} \, \mathrm{P}(\vec{n})  + e^{-\mu} \, \mathrm{P}(\vec{n}) \bigr] = 
2 \kappa \cosh\!  \mu \, \sum_{\vec{n}} \mathrm{Re} \bigl[ \mathrm{P}(\vec{n}) \bigr]  + 
i 2 \kappa \sinh \! \mu \, \sum_{\vec{n}} \mathrm{Im} \bigl[ \mathrm{P} (\vec{n}) \bigr] ~ .
\end{equation}
The first term of (\ref{eq:Action-imaginary}) and the first term of (\ref{action})
are combined into the real part of the action such that we find,
\begin{equation}
\begin{split}
S_R[\mathrm{P}] & = - \tau \sum\limits_{\vec{n} \in \Lambda} \sum\limits_{\vec{\nu}=1}^3 
\Bigl[ \mathrm{P}(\vec{n}) \, \mathrm{P}(\vec{n}+\vec{\nu})^{\dagger} + c.c. \Bigr] - 
2 \kappa \cosh \! \mu  \sum_{\vec{n}} \mathrm{Re} \bigl[ \mathrm{P}(\vec{n}) \bigr] \; , \; \\
S_I[\mathrm{P}] & = - 2 \kappa \sinh \! \mu \,  \mathrm{X[P]}  \; ,
\end{split}
\end{equation}
where we introduced,
\begin{equation}
\mathrm{X}[\mathrm{P}] = \sum_{\vec{n}} \mathrm{Im} \bigl[ \mathrm{P} (\vec{n}) \bigr] = \sum_{\vec{n}} \bigl[ \sin( \theta_1(\vec{n}) ) +  \sin( \theta_2(\vec{n}) ) -  \sin( \theta_1(\vec{n}) + \theta_2(\vec{n}) ) \bigr] ~.
\end{equation}
We now explore symmetries of the system to further simplify the partition sum and expectation values.
Charge conjugation ${\cal C}$ corresponds to the transformation $\forall \vec{n}$ :
$\theta_i(\vec{n}) \rightarrow - \theta_i(\vec{n})$, $i = 1,2$. 
Real and imaginary part of the action, and the reduced Haar measure transform as,
\begin{equation}
S_{R}[\mathrm{P}]  \overset{{\cal C}}{\longrightarrow} S_{R}[\mathrm{P}] \quad , \quad 
S_I[\mathrm{P}]  \overset{{\cal C}}{\longrightarrow} -S_I[\mathrm{P}] \quad , \quad 
d \mathrm{P}(\vec{n})  \overset{{\cal C}}{\longrightarrow} d \mathrm{P}(\vec{n}) \; ,
\end{equation}
such that the partition sum transforms as
\begin{equation}
\label{Ztrafo}
Z  = \int \!\! \mathcal{D}[\mathrm{P}] \; e^{-S_{R}[\mathrm{P}] - i S_I[\mathrm{P}] } \; 
\overset{{\cal C}}{\longrightarrow} \; \int \!\! \mathcal{D}[\mathrm{P}] \; e^{-S_{R}[\mathrm{P}] + i S_I[\mathrm{P}] } \; .
\end{equation}
Thus we can write the partition function as
\begin{equation}
Z  = \int \!\! \mathcal{D}[\mathrm{P}] \; e^{-S_{R}[\mathrm{P}] }  \cos( S_I[\mathrm{P}] ) = 
\int \!\! \mathcal{D}[\mathrm{P}] \; e^{-S_{R}[\mathrm{P}] }  \cos( 2 \kappa \sinh \! \mu \, \mathrm{X[P]} ) ~.
\end{equation}
Based on this form of the partition sum we now define the weighted density $\rho$, 
where we use as weight the Boltzmann factor with the real part of the action,
\begin{equation}
\rho (x) =  \int \!\! \mathcal{D} [\mathrm{P}] \; e^{-S_{R}[ \mathrm{P} ]} \, \delta(x-\mathrm{X[P]} ) 
\quad \text{where} \quad x \in [-x_{max},x_{max}] \; .
\label{rhodef}
\end{equation}
Here we have already used, that $\mathrm{X[P]}$ is bounded such that $x$ is restricted to the interval 
$[-x_{max},x_{max}]$ with $x_{max} = \frac{3\sqrt{3}}{2} V$, where $V$ denotes the number of sites of or lattice. 
Using again the transformation 
properties under charge conjugation one easily finds that the density is an even function of $x$, i.e., $\rho(x)=\rho(-x)$. 
Therefore, the partition function expressed in terms of the density is given by
\begin{equation}
\label{zrho}
Z \; = \; 2 \int\limits_{0}^{x_{max}} \!\! dx \, \rho(x) \cos( 2 \kappa \sinh \! \mu \, x)  \; .
\end{equation}
Expectation values of observables $\mathcal{O}(\mathrm{X[P]})$ which are a function of $\mathrm{X[P]}$ are given by
\begin{equation}
\label{vevrho}
\langle \mathcal{O} \rangle \; = \; \frac{2}{Z} \int\limits_{0}^{x_{max}} \!\! dx \, \rho(x) \, \Bigl[ \mathcal{O}_{E}(x) \, 
\cos( 2 \kappa \sinh\! \mu \, x) + i \, \mathcal{O}_{O}(x) \, \sin( 2 \kappa \sinh \! \mu \, x) \Bigr] ~,
\end{equation}
where we again used the fact that $\rho(x)$ is an even function and
$\mathcal{O}_{E}(x) = [\mathcal{O}(x) + \mathcal{O}(-x)]/2$ and 
$\mathcal{O}_{O}(x) = [\mathcal{O}(x) -\mathcal{O}(-x)]/2$ 
denote the even and odd parts of $\mathcal{O}(x) = \mathcal{O}_{E}(x) +\mathcal{O}_{O}(x) $.

The partition function (\ref{zrho}) and the expectation values (\ref{vevrho}) are obtained by integrating the density 
$\rho(x)$ with the factors $\cos( 2 \kappa \sinh\!  \mu \, x)$ and $\sin( 2 \kappa \sinh\! \mu \, x)$. 
While the density $\rho(x)$ is strictly positive, these factor are oscillating with $x$ and the frequency of the
oscillation increases exponentially with the chemical potential $\mu$  and linearly with the parameter $\kappa$. 
So for larger values of $\mu$ and $\kappa$ the density has to be computed with very high accuracy. 
This is how the complex action problem manifests itself in the density of states approach. 

\section{Computing the density of states with the FFA}

For the numerical evaluation we need to parameterize the density $\rho(x)$ in a suitable form. For that purpose we  
divide the interval $[0, x_{max}]$ into $N$ intervals $[x_n, x_{n+1}]$, 
$n=0,1, ... \, N\!-\!1$ with $x_0 = 0$ and $x_N = x_{max}$. We stress that the intervals can differ in their size 
$\Delta_n = x_{n+1} - x_n$, but clearly the sum rule $x_n = \sum_{j=0}^{n-1} \Delta_j$ must hold. It will turn out that 
working with variable interval sizes allows one to use finer resolutions for the density in regions of $x$ where 
$\rho(x)$ shows strong variations.

We now parameterize the density as 
\begin{equation}
\rho(x) \; = \; e^{-L(x)} \; ,
\label{rhoparam}
\end{equation}
where $L(x)$ is a continuous function which is piecewise linear, 
i.e., a straight line on each interval $[x_n, x_{n+1}]$. We stress at this point that the parameterization 
in the form (\ref{rhoparam}) with a piecewise linear function in the exponent is only an approximation, 
since for continuous degrees of freedom $\rho(x)$ cannot be represented with a finite number of parameters,
which here are given by the slopes in the intervals\footnote{This is different for applications in models
with discrete degrees of freedom, where a piecewise constant density provides an exact representation (see, e.g., 
\cite{Mercado:2014dva,Gattringer:2015lra}).}. Thus an exact representation can be obtained only in the limit 
$\Delta_n \rightarrow 0$, and below we will discuss a strategy how to determine an optimal choice for suitable 
interval sizes $\Delta_n$.

It is an interesting question how the interval sizes $\Delta_n$ affect the vacuum expectation values of observables. 
Obviously the $\Delta_n$ enter vacuum expectation values in a highly non-linear way and a thorough theoretical analysis of 
the effect of a finite discretization is beyond the scope of this study and has to be postponed to future work. However, 
in Section 5 we demonstrate numerically 
that already with moderate $\Delta_n$ we can reproduce the reference data from the dual simulation 
for a rather large range of chemical potential values. 

Denoting the slopes of the straight line on the interval $[x_n, x_{n+1}]$ by $k_n$, one can work out the 
explicit representation of $\rho(x)$,
\begin{equation}
\rho(x) \; = \; e^{-L(x)} \; = \; e^{- \sum\limits_{j=0}^{n-1} \Delta_j (k_j-k_n) - k_n x} \; 
= \; A_n \, e^{-k_n x} \quad \mbox{for} 
\quad x \in [x_n;x_{n+1}]  \; ,
\label{rhoparameterized}
\end{equation}
where we have fixed the (irrelevant) normalization of the density by setting $\rho(0)=1$, i.e., $L(0)=0$.
It is obvious, that in our parameterization the density depends only on the slopes $k_n$ of the piecewise
linear function $L(x)$. For later use we have introduced the notation $\rho(x) = A_n \, e^{- k_n x}$ in the second
equality of (\ref{rhoparameterized}), i.e., we have collected all factors independent of $x$ into the constant $A_n$.

To determine the density as given in (\ref{rhoparameterized}) we need to find the slopes $k_n$. For the
calculation of the $k_n$ we use so-called restricted vacuum expectation values 
$\langle \langle \mathcal{O} \rangle \rangle_{n}(\lambda)$, $n=0, ... \, N\!-\!1$, 
which depend on a free parameter $\lambda \in \mathds{R}$. They are defined as
\begin{equation}
\label{eq:restricted-value}
\begin{split}
\langle \langle \mathcal{O} \rangle \rangle_{n}(\lambda) & = \frac{1}{Z_{n}(\lambda)} 
\int \!  \mathcal{D}[\mathrm{P}] \, e^{-S_{R}[\mathrm{P}]+ \lambda \, X [\mathrm{P}] } \, 
\mathcal{O}\bigl( X [\mathrm{P}] \bigr) \, \theta_n\bigl( X[\mathrm{P}] \bigr) \; , \\
 Z_{n}(\lambda) & = \int \! \mathcal{D}[\mathrm{P}] \, 
 e^{-S_{R}[\mathrm{P}]+ \lambda \, X[\mathrm{P}] } \, \theta_n\bigl( X[\mathrm{P}] \bigr) \; ,
\end{split}
\end{equation}
with
\begin{equation}
\theta_n\bigl( x \bigr) \;  = \; \begin{cases}
1 \text{ for } x \in [x_n,x_{n+1}] \\
0 \text{ otherwise}
\end{cases} \; .
\end{equation}
The support function $\theta_n(x)$ restricts the values of $X[\mathrm{P}]$ in $Z_n(\lambda)$  and 
$\langle \langle \mathcal{O} \rangle \rangle_{n}( \lambda )$ to the interval $[x_n,x_{n+1}]$. Furthermore
we have also introduced an additional Boltzmann factor $e^{ \lambda \, X [\mathrm{P}] }$, which is real and positive, 
such that the generalized vacuum expectation values $\langle \langle \mathcal{O} \rangle \rangle_{n}(\lambda)$
can be evaluated with standard Monte Carlo techniques. The additional Boltzmann factor plays a twofold role:
By varying $\lambda$ we can systematically explore the density in all of the interval  $[x_n,x_{n+1}]$. 
In addition we can evaluate explicitly the dependence of $Z_{n}(\lambda)$ on the parameter  $\lambda$ and in
this way determine the $k_n$. This second role will be outlined now.

Writing $Z_{n}(\lambda)$ with the density and using the fact that on the interval $[x_n,x_{n+1}]$ the density is given by
$\rho(x) = A_n e^{-k_n x}$ (see (\ref{rhoparameterized})) we find,
\begin{equation}
\begin{split}
Z_{n}(\lambda) & \; =  \; \int \!\! dx \, \rho(x) \theta_n(x) e^{\lambda x} \; = \; 
\int_{x_n}^{x_{n+1}} \!\!\!\!  dx \; \rho(x) e^{\lambda x} \\
& \; = \; A_n \int_{x_n}^{x_{n+1}} \!\!\!\! dx \; e^{(\lambda - k_n) x} \; = \;   
A_n \, \frac{ e^{(\lambda - k_n) x_{n+1}} - e^{(\lambda - k_n) x_{n} } }{\lambda - k_n} \; .
\end{split}
\end{equation}
A simple calculation then gives (use $x_n = \sum_{j=0}^{n-1} \Delta_j$),
\begin{equation}
\langle \langle \mathrm{X[P]} \rangle \rangle_{n}(\lambda) \; = \;  
\frac{\partial}{\partial {\lambda} } \, \ln  Z_{n}(\lambda)  \; = \;  
\sum\limits^{n-1}_{j=0} \Delta_j + \frac{\Delta_n}{1-e^{-(\lambda-k_n)\Delta_n}} - \frac{1}{(\lambda-k_n)} \; ,
\end{equation}
which we rearrange to new, rescaled observables $Y_n(\lambda)$ defined as,
\begin{equation}
Y_n(\lambda) \equiv 
\frac{\langle \langle \mathrm{X[P]} 
\rangle \rangle_{n}(\lambda) - \!\!\sum^{n-1}_{j=0} \!\! \Delta_j }{\Delta_n} - \frac{1}{2} =   
\frac{1}{1\!-\!e^{-(\lambda-k_n)\Delta_n}} - \frac{1}{(\lambda\!-\!k_n)\Delta_n} -\frac{1}{2}  \
\equiv  h\Big(\!(\lambda-k_n)\Delta_n \!\Big)  , 
\label{eq:sigmoid-function}
\end{equation}
where in the last step we introduced the abbreviation $h(s) = 1/(1-e^{-s}) - 1/s - 1/2$.

Using Eq.~(\ref{eq:sigmoid-function}) we can now determine the slopes $k_n$ 
in a simple way: On the left hand side we have the observable $Y_n(\lambda)$ which is related 
to the restricted vacuum expectation value $\langle \langle \mathrm{X[P]} \rangle \rangle_{n}(\lambda)$ 
which can be evaluated with standard Monte Carlo techniques for several values of $\lambda$. 
These Monte Carlo results are described by $h \Big(\!(\lambda-k_n)\Delta_n \!\Big)$ where $h(s)$ is a simple
smooth function. Thus on the right hand side of Eq.~(\ref{eq:sigmoid-function}) we have a single free parameter,
namely the slope $k_n$ which we want to determine. Its value is then obtained by a simple one-parameter fit of 
$h \Big(\!(\lambda-k_n)\Delta_n \!\Big)$ to the numerical data. This procedure is the reason for referring to our
method as the ''Functional Fit Approach'' (FFA). Since we determine the $k_n$ from a fit of the data for $Y_n(\lambda)$, 
the normal $1/\sqrt{N}$ dependence of the errors on the statistics $N$ applies.

\begin{figure}[t]
\begin{center}
\includegraphics[width=0.95\textwidth]{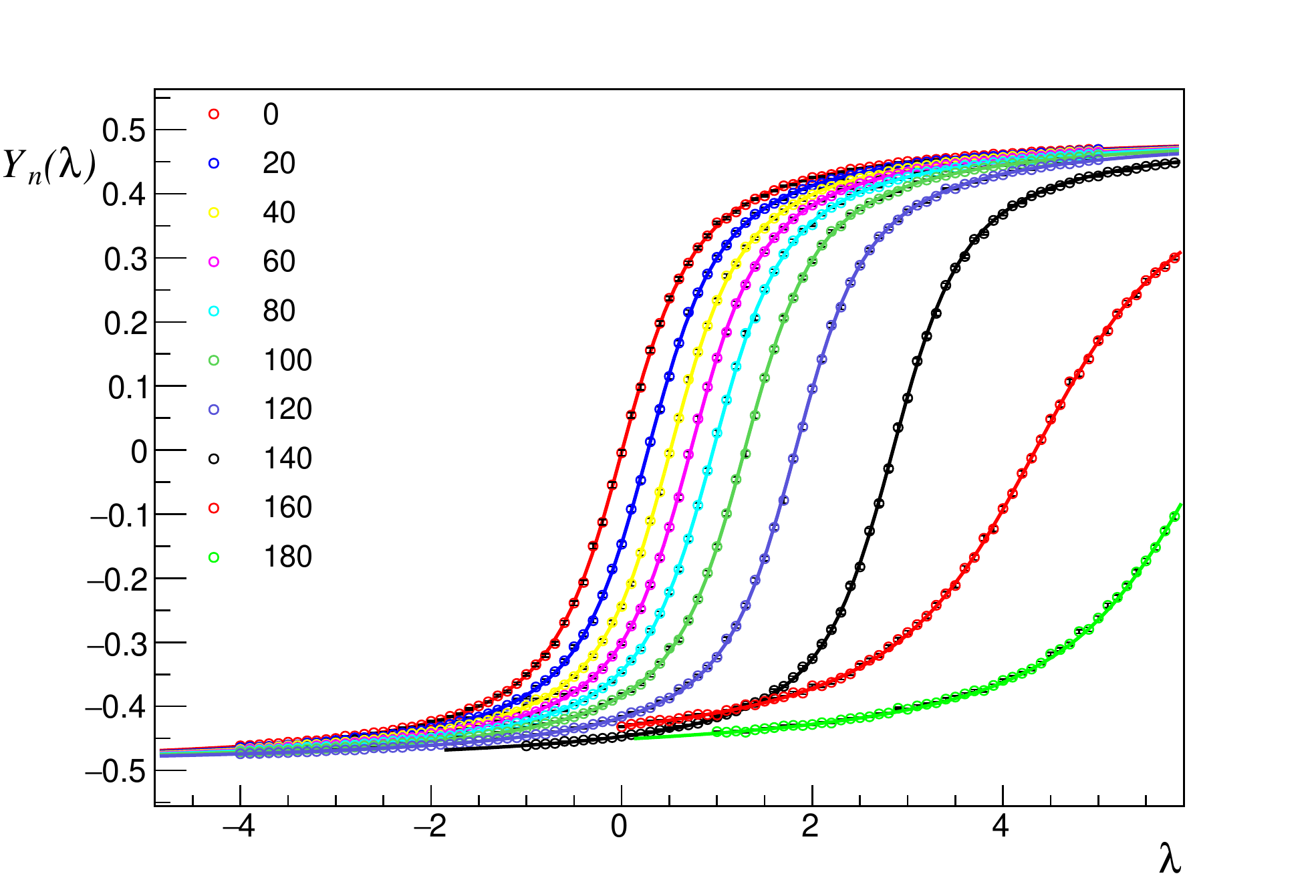}
\end{center}
\caption{Monte Carlo results for $Y_{n}(\lambda)$ as defined in Eq.~(\protect{\ref{eq:sigmoid-function}}) for a 
$8^4$ lattice with $\tau=0.075$, $\kappa = 0.005$ and $\mu=0.0$. We show the  results for $Y_{n}(\lambda)$ as a 
function of $\lambda$ and compare different values of $n$ between $n = 0$ and $n = 180$ (circles of different color;  
the data sets move to the right with increasing $n$). The full curves are the results of the one-parameter 
fits with the function $h\Big(\!(\lambda-k_n)\Delta_n \!\Big)$ on the right hand side of 
Eq.~(\protect{\ref{eq:sigmoid-function}}).
\label{fig:fit-example}}
\end{figure}

It is straightforward to show that $h(s)$ is monotonically increasing with $h(0) = 0, h^\prime(0) = 1/12$ and 
$\lim_{s \rightarrow \pm \infty} h(s) = \pm 1/2$. Thus the function $h\Big(\!(\lambda-k_n)\Delta_n \!\Big)$ used for the fit
has a single zero at $\lambda = k_n$ where it has a slope of $\Delta_n/12$. In principle one could determine
the zero of $h\Big(\!(\lambda-k_n)\Delta_n \!\Big)$ to find $k_n$\footnote{This is the approach chosen in the LLR 
method \cite{Langfeld:2012ah,Langfeld:2013xbf,Langfeld:2014nta,Langfeld:2015fua,Garron:2016noc}, where 
one essentially searches for the zero of $h\Big(\!(\lambda-k_n)\Delta_n \!\Big)$ using an iterative
process.}, but using all values of $\lambda$ in a fit with 
$h\Big(\!(\lambda-k_n)\Delta_n \!\Big)$ makes use of all generated Monte Carlo data. 
A second important aspect is that the quality of the fit also allows for a consistency check of the method: If one finds
that the Monte Carlo data are not well described by the function $h\Big(\!(\lambda-k_n)\Delta_n \!\Big)$ this is an
indication that the interval size $\Delta_n$ was chosen too large. We remark again, that we here work 
with variable interval sizes $\Delta_n = x_{n+1} - x_n$ and in regions of $x$ where the density $\rho(x)$ 
shows a large variation one can choose smaller interval sizes $\Delta_n$.

\begin{figure}[t]
\centering
\hspace*{-3mm}
\subfloat[][\emph{Density as a function of $x$}.]
{\includegraphics[width=0.52\textwidth]{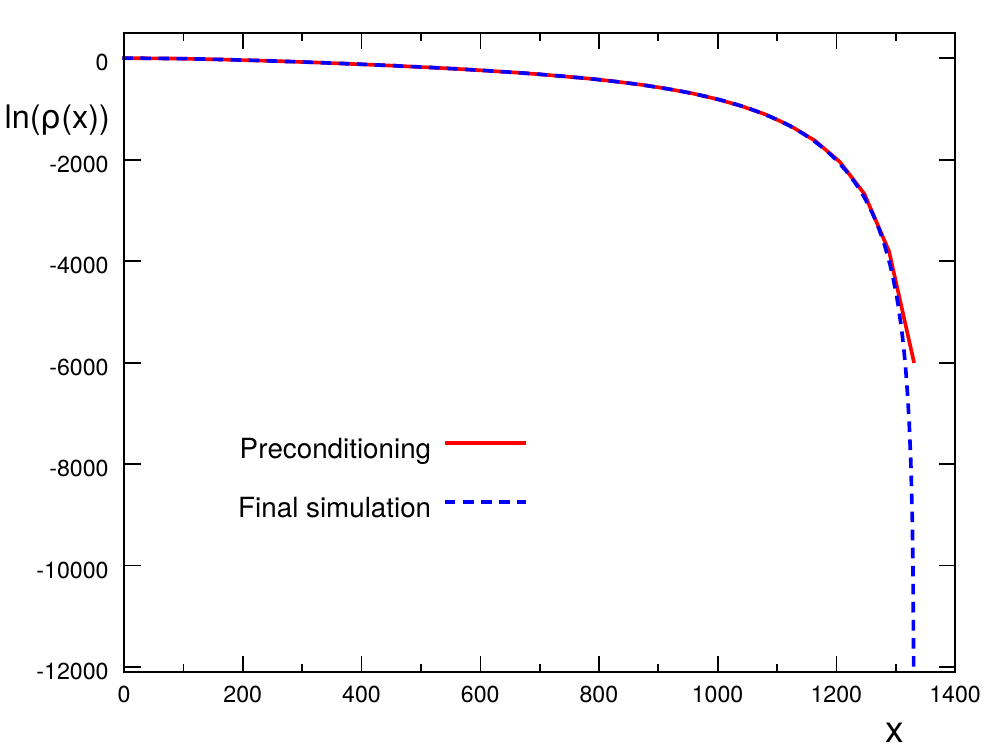}} 
\subfloat[][\emph{Zoom for small values of $x$}.]
{\includegraphics[width=0.52\textwidth]{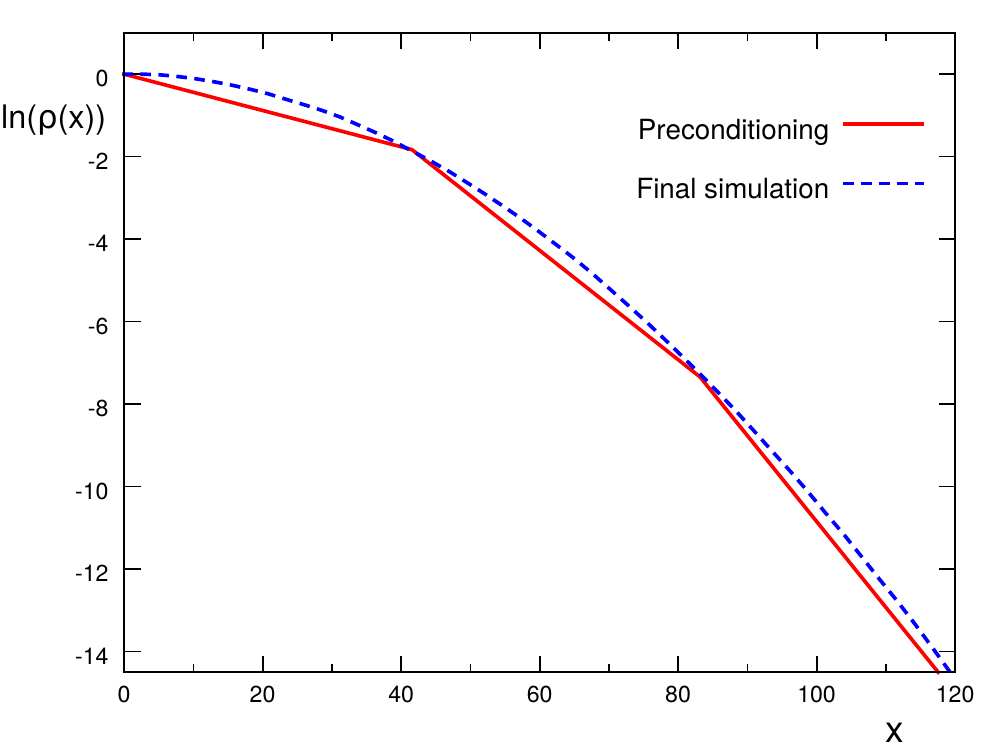}} 
\caption{Comparison of the density from the final simulation (dashed curve) and the results from preconditioning 
(full lines). The data are for $8^3$ lattices at $\tau = 0.101, \kappa = 0.04$ and $\mu = 1.0$.
We find that preconditioning describes well the overall shape but the zoom on the right hand side shows
that the preconditioning density is too coarse for evaluating observables.}
\label{fig:preconditioning}
\end{figure}

\begin{figure}[t]
\begin{center}
\includegraphics[width=0.69\textwidth]{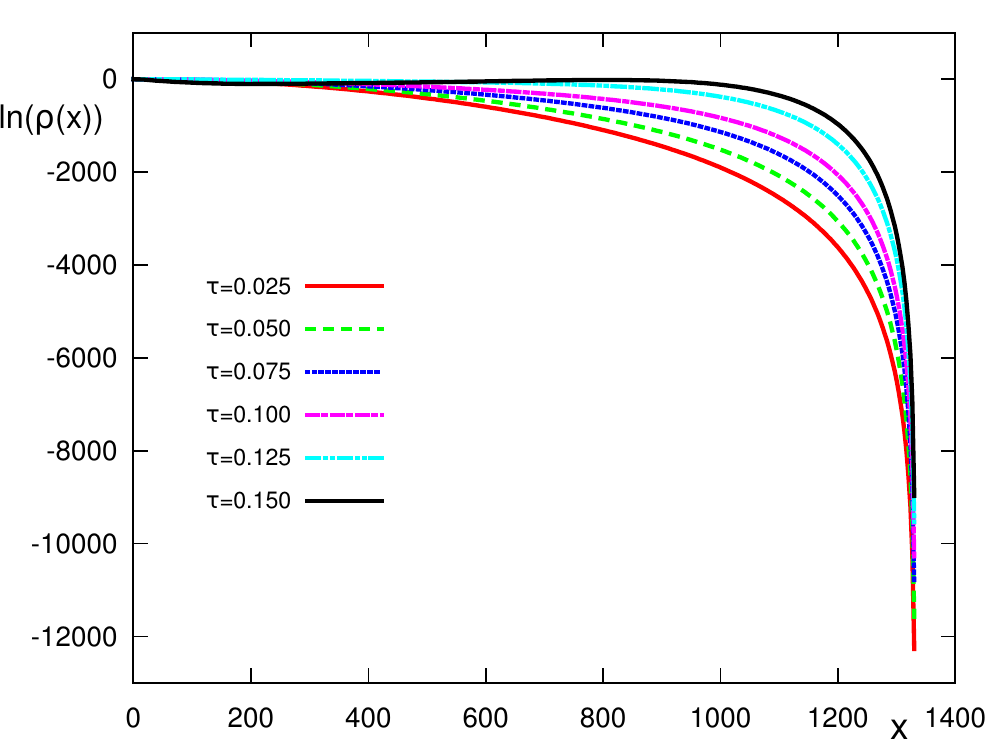}
\end{center}
\caption{Comparison of the logarithm of the density of states $\rho(x)$ 
for different values of $\tau$. The data are for lattice size $8^3$ at fixed $\kappa=0.005$ and fixed $\mu=1.0$.}
\label{fig:rho-tau}
\end{figure}

In Fig.~\ref{fig:fit-example} we show an example of the fit procedure on a small lattice. The circles are the results of the
evaluation of $Y_n(\lambda)$ with the restricted Monte Carlo simulations. We show the results for several values of $n$ 
between $n = 0$ and $n = 180$ (circles of different colors; the data sets move to the right with increasing $n$). 
The full curves represent the functions 
$h\Big(\!(\lambda-k_n)\Delta_n \!\Big)$ where the $k_n$ were determined from a fit of the Monte Carlo data. It is
obvious that the Monte Carlo data are very well represented by the functions $h\Big(\!(\lambda-k_n)\Delta_n \!\Big)$. 
Note that for the largest two $n$ we show ($n = 160$ and $n = 180$) a smaller $\Delta_n$ was used, 
which leads to a smaller slope $\Delta_n/12$ at the point where $h\Big(\!(\lambda-k_n)\Delta_n \!\Big)$ crosses zero.
The reason for choosing a smaller $\Delta_n$ is the fact that for large $x$ the density varies faster such that 
a finer resolution, i.e., finer intervals $[x_n, x_{n+1}]$ are advisable. Once the $k_n$ are obtained from the fits
one can determine the density $\rho(x)$ from the explicit expression (\ref{rhoparameterized}). Examples for the density
are shown in Figs.~\ref{fig:preconditioning}, \ref{fig:rho-tau} and \ref{fig:rho-change-zoom}.

Let us finally comment on the choice of the values $\lambda$ where the simulations for 
determining $Y_n(\lambda)$ are done. 
In Fig.~\ref{fig:fit-example} we observe that for larger $n$ the curves shift to the right,
which is due to the fact that for larger values of $x$, which corresponds to intervals $[x_n, x_{n+1}]$ with larger $n$, 
the density $\rho(x)$ has larger slopes $k_n$. To obtain optimal fit results one should have Monte Carlo data
in a range of $\lambda$ that properly brackets the value $\lambda = k_n$ where the fit function  
$h\Big(\!(\lambda-k_n)\Delta_n \!\Big)$ has its zero. 

To avoid exploring a large range of values for $\lambda$ we developed a strategy which we refer to as 
''pre-conditioning''. The idea is to do a first run with only a small number $N$ of intervals, which then of course 
have rather large sizes $\Delta_n$ and give rise to only a crude approximation of $\rho(x)$. 
However, analyzing this crude
approximation we already have an idea what size the slopes $k_n$ are. This information about the $k_n$ 
can then be used for:
\begin{enumerate}
\item Determining the optimal interval sizes $\Delta_n$ (small for regions with a large variation of $\rho(x)$ and 
larger for regions with little change of $\rho(x)$).  

\item Identifying suitable ranges for the values of $\lambda$ in the 
determination of $Y_n(\lambda)$ such that the corresponding value $\lambda = k_n$ is properly bracketed by the
Monte Carlo data.  
\end{enumerate}

In Fig.~\ref{fig:preconditioning} we illustrate preconditioning and compare the final results for
the density to a coarse approximation with only few and large intervals. The two plots (the rhs.\ is a zoom into the
small-$x$ range) show that the preconditioning results already give a good first estimate of the final density which 
can be used to optimize the runs for the target resolution with fine intervals $[x_n, x_{n+1}]$. On the other hand it is 
clearly seen in the zoom on the right hand side that the coarse density obtained in the preconditioning step is not suitable 
for computing observables with the fluctuating integrals of (\ref{vevrho}).

We remark at this point that the preconditioning idea can be pushed further: One can include the coarse density obtained
from a preconditioning calculation into the definition of the density such that the final density is obtained as a correction 
to the preconditioning density. It is relatively simple, although somewhat technical, to work out the corresponding 
equations along the lines of Eqs.~(\ref{rhodef}) -- (\ref{eq:sigmoid-function}). One finds that the final density
has much smaller values of $k_n$ since they are only corrections relative to the slopes of the preconditioning
density. Thus the corresponding relevant ranges of $\lambda$ are all centered near $\lambda \sim 0$ and the 
effort for the determination of the optimal range for different $n$ is considerably reduced. 

It is clear that the shape of the density $\rho(x)$ will be different for different parameter values. For a first illustration 
in Fig.~\ref{fig:rho-tau} we show $\rho(x)$ for fixed values of $\mu$ and $\kappa$ but different values of
$\tau$. For a comparison of $\rho(x)$ at different values of $\mu$ see Fig.~\ref{fig:rho-change-zoom} below.

\section{Details of the restricted Monte Carlo simulations}

Before we come to a more detailed discussion of our results, in this section we provide a short overview over the 
technical aspects of our simulations used in this paper, designed for assessing the DoS FFA method and to further 
develop this approach.

The DoS FFA method is based on fitting the Monte Carlo data for $Y_n(\lambda)$ which is related to 
the restricted expectation value $\langle \langle \mathrm{X[P]} \rangle \rangle_{n}(\lambda)$ as defined in 
(\ref{eq:restricted-value}).  Since the restricted vacuum expectation value requires $\mathrm{X[P]} \in [x_n, x_{n+1}]$, 
we first need to generate an initial configuration of the angles $\theta_1(\vec{n})$ and $\theta_2(\vec{n})$
such that this constraint is obeyed. Such a configuration with constant spin values $\mathrm{P}(\vec{n})$ for all 
$\vec{n} \in \Lambda$ can easily be constructed by hand, but of course needs to be equilibrated before taking 
measurements. For this and the subsequent computation of observables the restricted vacuum expectation values 
require a slightly modified Monte Carlo update. It contains an additional step which rejects proposal 
configurations $\mathrm{P}$ that violate the condition $\mathrm{X[P]} \in [x_n, x_{n+1}]$. 
In the simulation of our SU(3) spin system we did not observe problems due to this additional rejection step 
and the acceptance rate remained reasonably high for all parameter values we tested. As we will see, with the 
parameters chosen here we can cover a rather large range of chemical potential values, but is is also clear that for 
pushing to even higher values one would have to use considerably smaller intervals and the additional rejection 
step might slow down the algorithm.  

In addition to running the simulations for all $N$ intervals, in each interval $[x_n, x_{n+1}], n = 0, 1 \, ... \, N-1$ 
we have to produce data for a suitably chosen set of $\lambda$ values such that we can perform 
an optimal fit of $Y_n(\lambda)$ with $h\Big(\!(\lambda-k_n)\Delta_n \!\Big)$. Thus the approach requires a total
of $N \times N_\lambda$ restricted Monte Carlo simulations, where $N_\lambda$ is the number of $\lambda$ values 
used for the fit. The number $N$ of intervals of a given size $\Delta_n$ 
scales with the volume $V$ of the lattice since $\mathrm{X[P]}$ is an 
extensive quantity, while $N_\lambda$ is a fixed number independent of the system size.

In this exploratory study we show data for lattice volumes of $8^3$ and $12^3$. We use between $N = 200$ and 
$N = 600$ intervals and $N_{\lambda}\approx60-100$. We use $\mathcal{O}(10^4)$ configurations for evaluating each 
$\langle \langle \mathrm{X[P]} \rangle \rangle_{n}(\lambda)$ and these configurations are separated by 
10 sweeps for decorrelation.  All errors we display are statistical errors. For their determination 
we computed the error for each $\langle \langle \mathrm{X[P]} \rangle \rangle_{n}(\lambda)$ via the standard deviation, 
determined the error for the $k_n$ from the fit with $h\Big(\!(\lambda-k_n)\Delta_n \!\Big)$ and subsequently 
performed a Monte Carlo propagation of this error. This procedure allows to simultaneously take into account the
influence of the errors of all Monte Carlo data on the statistical error we show for the final observables. 

As outlined in the previous section we use a preconditioning phase where we perform a simulation with only few and 
large intervals and low statistics. Typically we use $N =  50$ to $N = 100$ intervals and $N_{\lambda}\approx20-30$ 
with a statistics of only $\mathcal{O}(10^2)$ configurations. However, this additional low-cost simulation 
considerably speeds up the final simulation since we can use suitably chosen interval sizes $\Delta_n$
and optimal ranges for the values of $\lambda$.

\section{Observables and their comparison to a dual simulation}

So far we have only discussed techniques how to compute the density $\rho(x)$ and have 
shown some of the corresponding results.
However, the true test of a density of states approach is of course the assessment of observables, since in their 
evaluation according to  (\ref{vevrho}) the density is integrated over with highly oscillating factors that probe the fine
details of $\rho(x)$. Thus only observables test if the determination of $\rho(x)$ is sufficiently accurate and delimit 
up to which values of $\mu$ the approach is reliable since, as we have discussed in Section 2, the frequency of 
oscillation increases exponentially with $\mu$. It would be very interesting and important to perform a systematic 
comparison of accuracy and computational cost of physical observables 
to the results from other modern DoS techniques, in particular the LLR method 
\cite{Langfeld:2012ah,Langfeld:2013xbf,Langfeld:2014nta,Langfeld:2015fua,Garron:2016noc}, where, however, 
the focus so far was on technical development and on the density itself.

As pointed out in the introduction, one of the main reasons for choosing the SU(3) spin model as a test case for
developing and assessing the DoS FFA method is the fact that the model has a dual representation 
\cite{Gattringer:2011gq} that solves the complex action problem. Dual Monte Carlo simulations in terms of worldlines
\cite{Mercado:2012ue,Delgado:2012uh} are possible for arbitrary $\mu$ and provide reference data for assessing the 
new DoS FFA techniques.

For the comparison to the dual simulation we use the particle number $n$ and the corresponding susceptibility $\chi_n$.
They are obtained as derivatives of $\ln Z$ and can be computed easily in the dual approach. Their exact definition and
the representation in terms of the density $\rho(x)$ are given by
\begin{eqnarray}
n & = & \frac{1}{V} \, \frac{1}{2\kappa} \, \frac{\partial}{\partial \sinh \! \mu } \, \ln Z 
\; = \; \frac{1}{V} \, \frac{2}{Z} \, \int\limits_{0}^{x_{max}} \! \! dx \, \rho(x) \, \sin( 2 \kappa \sinh\! \mu \, x ) \, x \; ,
\label{eq:obs_n} \\
\chi_{n} & = & \frac{1}{V} \, \frac{1}{(2\kappa)^2} \, \frac{\partial^2 }{\partial \sinh \! \mu^2  } \ln Z 
\label{eq:susce_n} \\ 
& = & \frac{1}{V} \, \left[ \frac{2}{Z} \int\limits_{0}^{x_{max}} \! \! dx \, \rho(x) \, \cos(2 \kappa \sinh \! \mu \, x) \, x^2 + 
\left( \frac{2}{Z} \int\limits_{0}^{x_{max}} \! \! dx \, \rho(x) \, \sin( 2 \kappa \sinh \! \mu \, x \right)^{\!2} \, \right]  .
\nonumber
\end{eqnarray}
In the dual formulation one can perform the derivatives of $\ln Z$ in the same way and obtains the observables as 
moments of dual variables \cite{Gattringer:2011gq,Mercado:2012ue,Delgado:2012uh}.

\begin{figure}[p]
\vspace*{-8mm}
\begin{center}
\hspace*{-3mm}
{\includegraphics[width=0.5\textwidth]{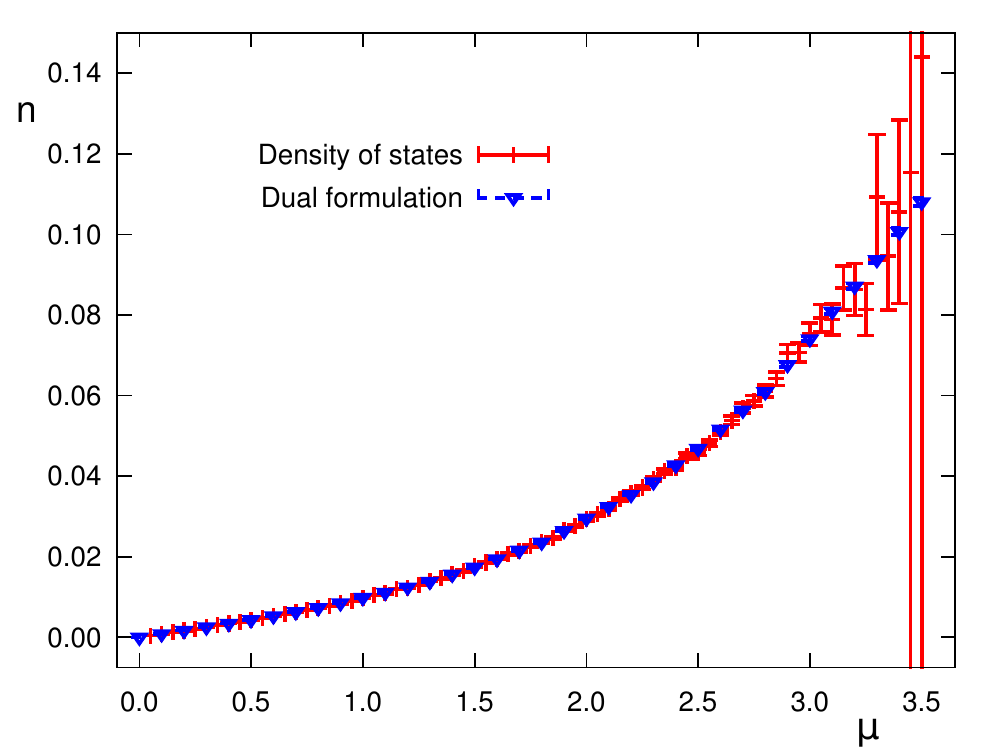}}  
{\includegraphics[width=0.5\textwidth]{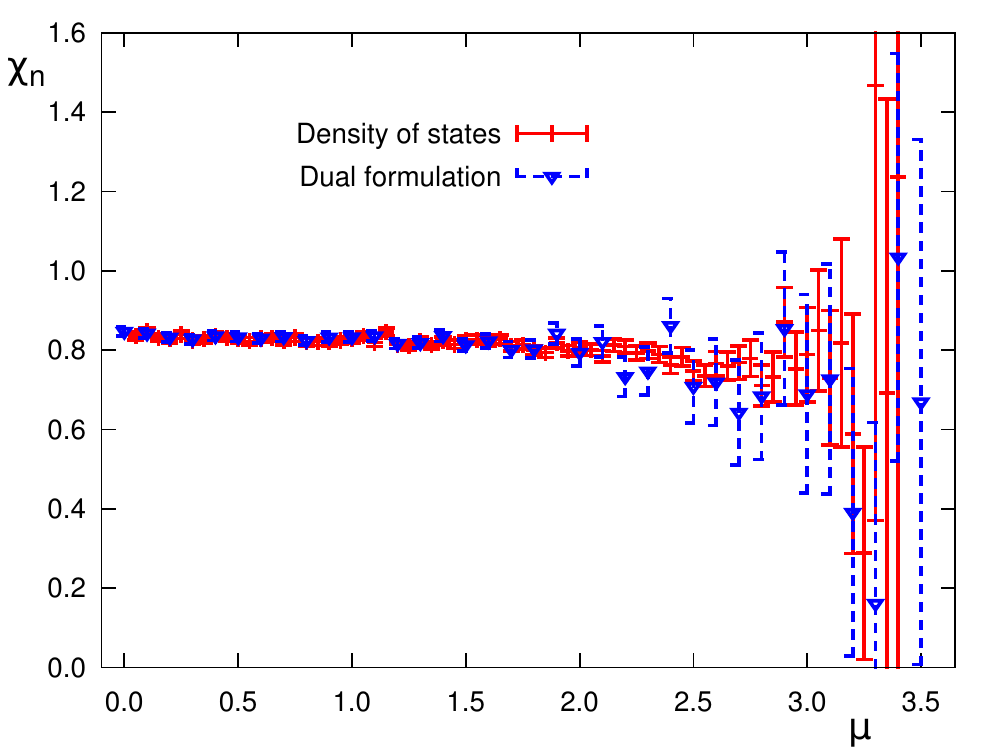}} 
\end{center}
\vspace*{-3mm}
\caption{$n$ (lhs.) and $\chi_n$ (rhs.) as function of $\mu$ ($8^3$ lattice with $\kappa=0.005$ and $\tau=0.066$).
We compare the DoS FFA results to the reference data from the dual formulation.}
\label{fig:results_tau0066}
\begin{center}
\hspace*{-3mm}
{\includegraphics[width=0.5\textwidth]{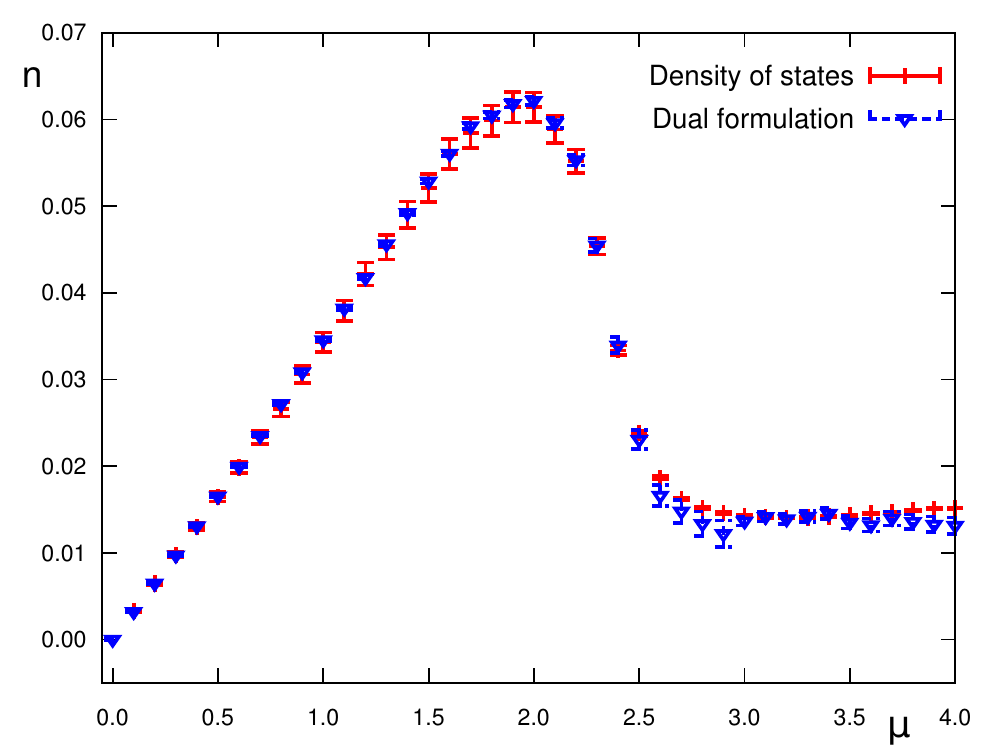}}  
{\includegraphics[width=0.5\textwidth]{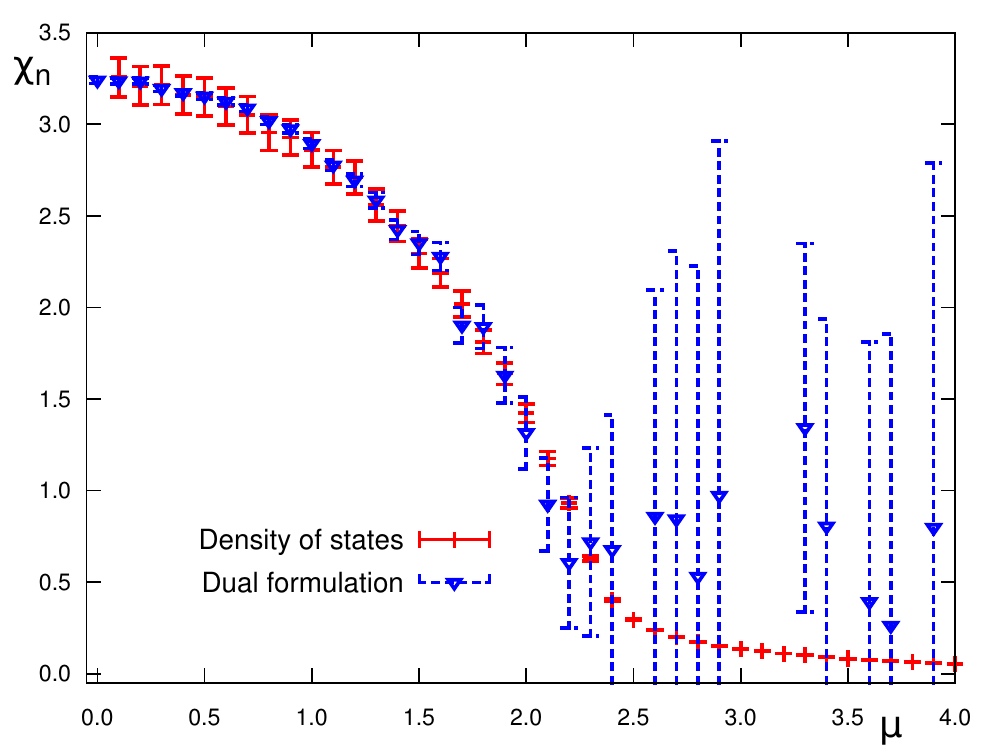}} 
\end{center}
\vspace*{-3mm}
\caption{$n$ (lhs.) and $\chi_n$ (rhs.) as function of $\mu$ ($8^3$ lattice with $\kappa=0.005$ and $\tau=0.13$).
We compare the DoS FFA results to the reference data from the dual formulation.}
\label{fig:results_8_murun}
\begin{center}
\hspace*{-3mm}
{\includegraphics[width=0.5\textwidth]{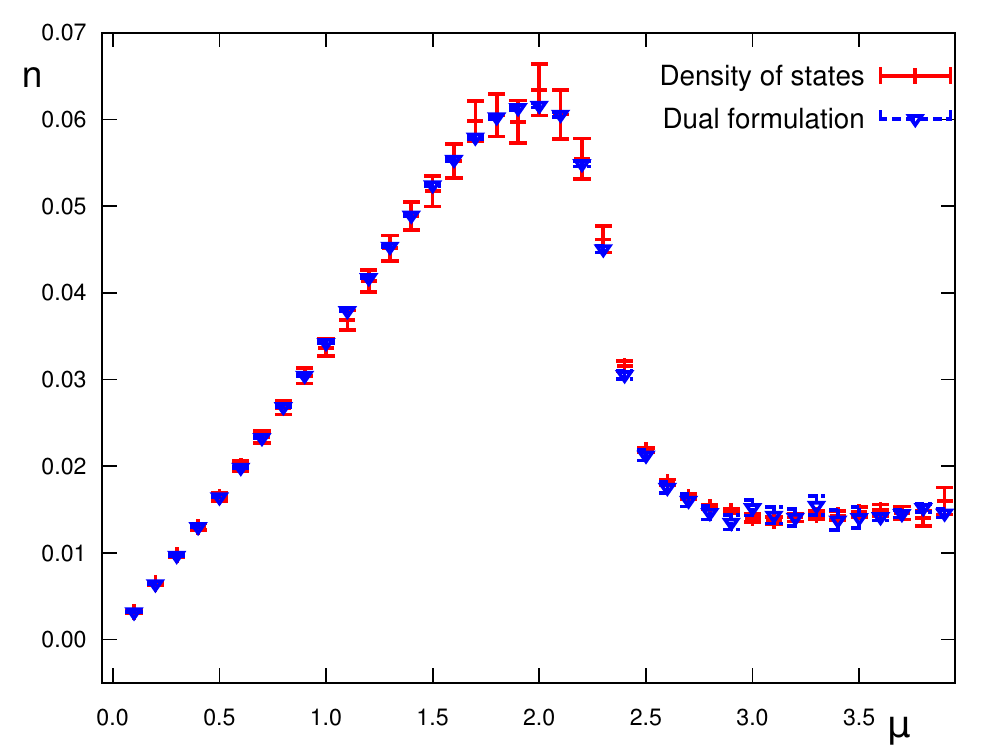}}
{\includegraphics[width=0.5\textwidth]{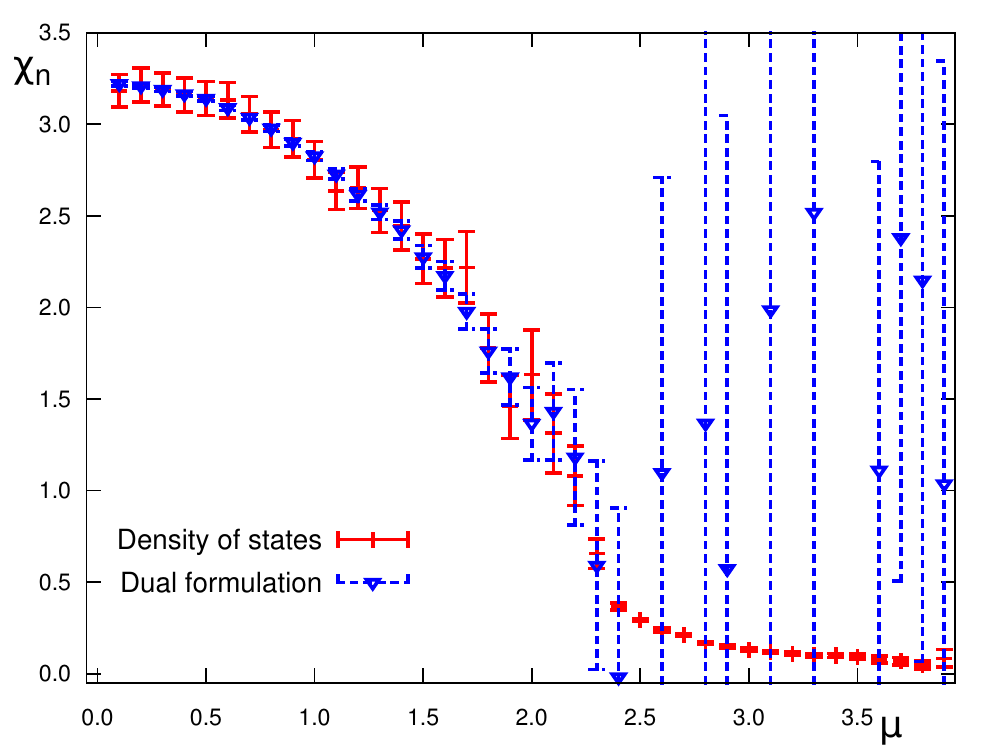}} 
\end{center}
\vspace*{-3mm}
\caption{$n$ (lhs.) and $\chi_n$ (rhs.) as function of $\mu$ ($12^3$ lattice with $\kappa=0.005$ and $\tau=0.13$).
We compare the DoS FFA results to the reference data from the dual formulation.}
\label{fig:results_12_murun}
\end{figure}

We begin the discussion of our results with Fig.~\ref{fig:results_tau0066}, where we show $n$ and $\chi_n$ as a function 
of $\mu$ on a $8^3$ lattice at $\kappa = 0.005$ and $\tau=0.066$. Here we use a relatively coarse discretization 
with $N = 450$ intervals and a low statistics of $5000$ measurements for each 
$\langle \langle \mathrm{X[P]} \rangle \rangle_{n}(\lambda)$. 
We find that with these parameters the density $n$ (lhs.\ plot) 
from DoS FFA agrees well with the dual results for chemical potentials up to $\mu \sim 3.0$, while for the corresponding 
susceptibility we find good agreement up to $\mu \sim 2.5$. Note that for the second moment also the dual results 
show quite large errors due to long autocorrelation times of the dual simulation in this region of the parameter space. 

An interesting question is how one should choose the overall scale for the interval size $\Delta_n$. We already 
discussed that in regions of $x$ with a large variation of $\rho(x)$ one chooses smaller $\Delta_n$, but we 
have not yet addressed the question what is a suitable choice for $\Delta_n$ in general. For 
computing observables according to Eq.~(\ref{vevrho}) we integrate the density with $\sin( 2\kappa \sinh\! \mu \, x)$ and
$\cos( 2\kappa \sinh\! \mu \, x)$. Thus, for a given $\kappa$ and $\mu$ one full oscillation corresponds to 
$\Delta x = 2\pi/ 2\kappa \sinh\! \mu$, which is the scale at which the fluctuating factors probe $\rho(x)$. 
Clearly the resolution of $\rho(x)$ given by the interval size $\Delta_n$ must be considerably smaller than 
$\Delta x$ and we obtain as criterion
\begin{equation}
\label{eq:thumb}
\Delta_n \ll \frac{2 \pi}{2 \kappa \sinh{\mu}} ~.
\end{equation}  

In Fig.~\ref{fig:results_8_murun} we show results where we used a much higher statistics of 
75000 measurements and chose intervals that are slightly smaller compared to those 
used in Fig.~\ref{fig:results_tau0066}. Furthermore we here switched to $\tau=0.130$ (keeping $\kappa = 0.005$),
a value where the system undergoes a crossover near $\mu = 2.0$ (see \cite{Mercado:2012ue}). 
Determining the observables across this crossover is clearly a more challenging task than the 
calculation at $\tau = 0.066$ in Fig.~\ref{fig:results_tau0066}, where there is no crossover in the range of
$\mu$ values considered  \cite{Mercado:2012ue}. Nevertheless, with the improved statistics and smaller 
$\Delta_n$ we find very good agreement with the dual formulation all the way up to $\mu = 4.0$ (although the 
dual data for $\chi_n$ again suffer from large errors due to autocorrelation at these parameter values). It is interesting to 
note that the agreement remains good across the crossover near $\mu \sim 2$, which is visible in a
maximum of $n$. Careful inspection of the error bars for the DoS FFA results show that they even become 
smaller again for $\mu > 2$, which is an unexpected behavior that needs 
some extra consideration which we present below. 
For completeness we also considered the observables on larger lattices, and for the same parameters as in 
Fig.~\ref{fig:results_8_murun}, i.e., $\kappa = 0.005$ and $\tau = 0.130$ we show our results for lattice size
$12^3$ in Fig.~\ref{fig:results_12_murun}. The behavior is almost exactly the same as on the smaller lattices 
used in Fig.~\ref{fig:results_8_murun}. 

\begin{figure}[t]
\centering
{\includegraphics[width=0.75\textwidth]{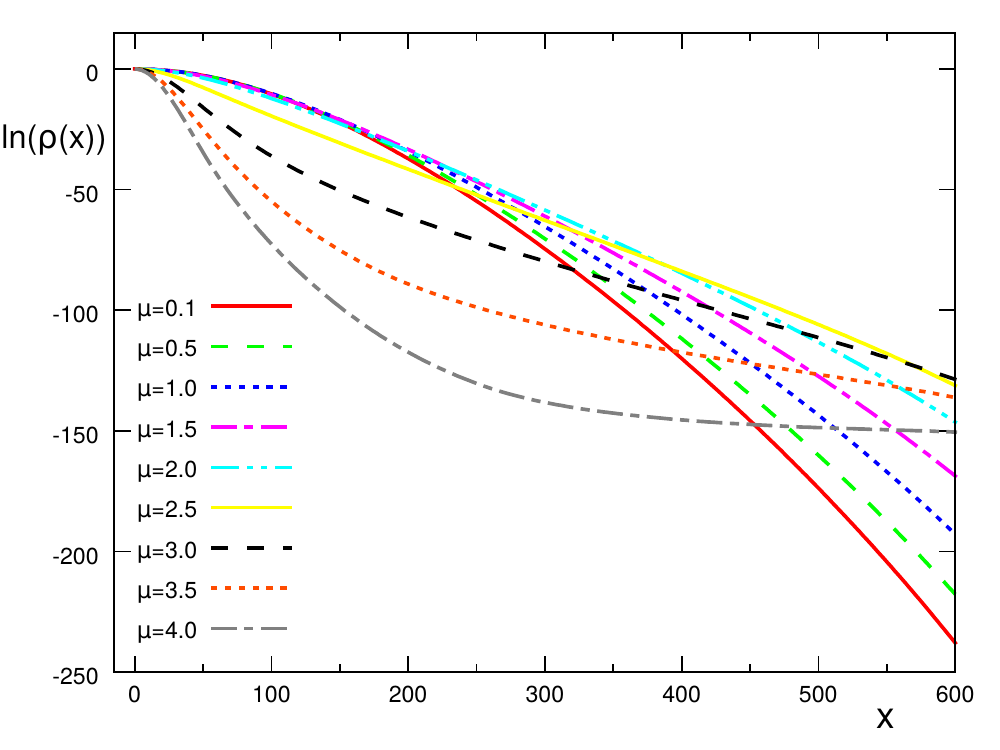}}
\caption{Comparison of  $\ln \rho(x)$ for different values of $\mu$. The data are for $8^3$ lattices and 
$\tau=0.130$ and $\kappa=0.005$, i.e., the parameters used in Fig.~\ref{fig:results_8_murun} (note that we show 
only the first half of the full range of $x$).}
\label{fig:rho-change-zoom}
\end{figure}

Let us come back to the interesting observation we made in 
Figs.~\ref{fig:results_8_murun} and \ref{fig:results_12_murun}: 
For $\mu > 2$ the error bars of the DoS data 
become smaller again. This is unexpected because we have pointed out that when 
increasing $\mu$ the frequency of the oscillating factors in (\ref{vevrho}) increases exponentially and the density 
is probed on even finer scales. The solution of this riddle is the fact that also the density $\rho(x)$ depends
on the chemical potential, since $\mu$ enters the real part of the action which is used for defining $\rho(x)$ in 
Eq.~(\ref{rhodef}). Indeed it turns out that when increasing $\mu$ above the crossover value $\mu \sim 2$ the 
shape of $\rho(x)$ changes considerably. This is illustrated in Fig.~\ref{fig:rho-change-zoom} where we 
show the density $\rho(x)$ at different values of $\mu$ for the parameter values used in Fig.~\ref{fig:results_8_murun}. 
It is obvious that the behavior changes considerable when increasing $\mu$ above 2, and the density decreases
much faster as a function of $x$ for larger values of $\mu$. This fast decrease suppresses contributions at 
large $x$ and thus leads to faster convergence in (\ref{vevrho}) and thus smaller errors. We conclude that the 
applicability, accuracy and convergence range in $\mu$ of DoS methods not only depends on the parameters 
$N$, $\Delta_n$ and the statistics used, but also on the underlying physics.

\begin{figure}[t]
\begin{center}
\hspace*{-3mm}
{\includegraphics[width=0.5\textwidth]{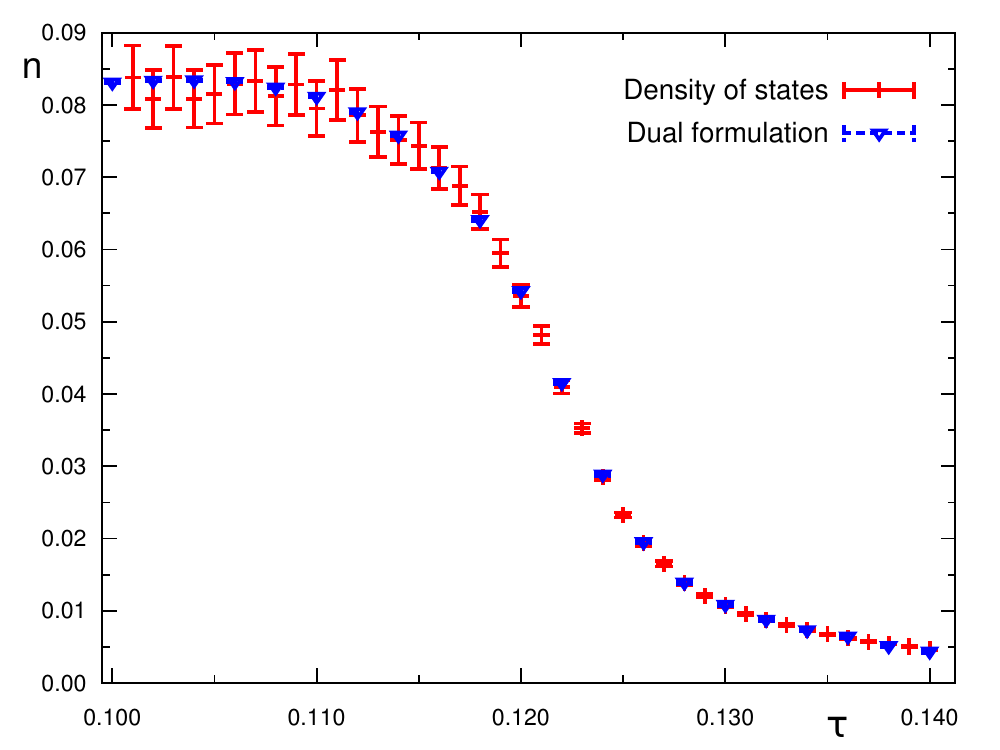}} 
{\includegraphics[width=0.5\textwidth]{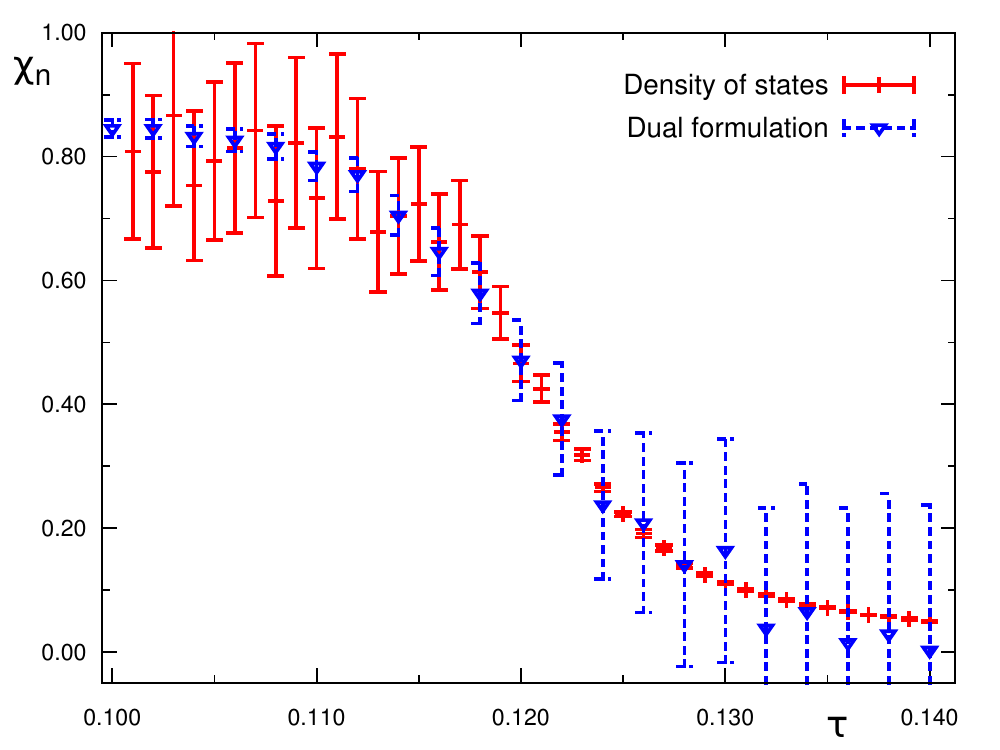}} 
\end{center}
\caption{$n$ (lhs.) and $\chi_n$ (rhs.) as function of $\tau$ ($8^3$ lattice with $\kappa=0.005$ and $\mu=1.0$).
We compare the DoS FFA results to the reference data from the dual formulation.}
\label{fig:results_8_taurun}
\end{figure}

We continue our discussion of observables with showing the results for a vertical section in the 
$\mu$-$\tau$ plane. In Fig.~\ref{fig:results_8_taurun} we work at fixed $\kappa = 0.005$ and fixed 
$\mu = 1.0$ and study the observables $n$ and $\chi_n$ as a function of $\tau$.  In this case 
one crosses a first order transition near $\tau = 0.13$ as determined from the Polyakov loop susceptibility in  
\cite{Mercado:2012ue}. Although the volume is rather small in our study, 
it is still remarkable that also here we find very good agreement between the DoS FFA results and the reference data
from the dual simulation. This indicates that DoS methods could indeed be competitive with other approaches also 
in the vicinity of phase transitions. 

\begin{figure}[t]
\begin{center}
\hspace*{-3mm}
{\includegraphics[width=0.5\textwidth]{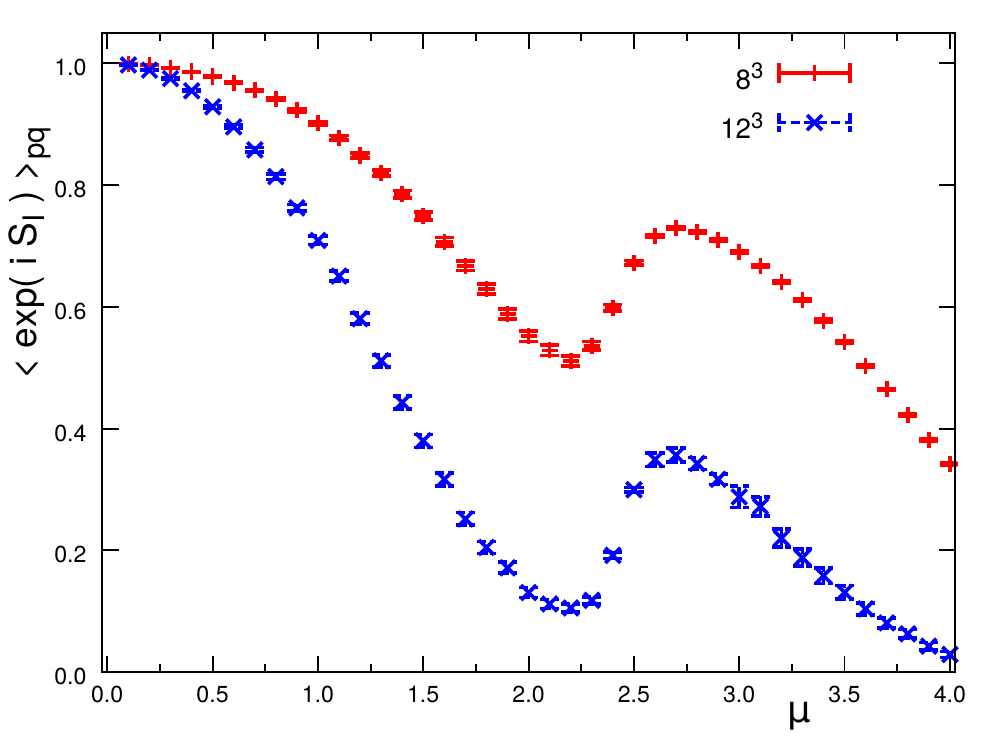}} 
{\includegraphics[width=0.5\textwidth]{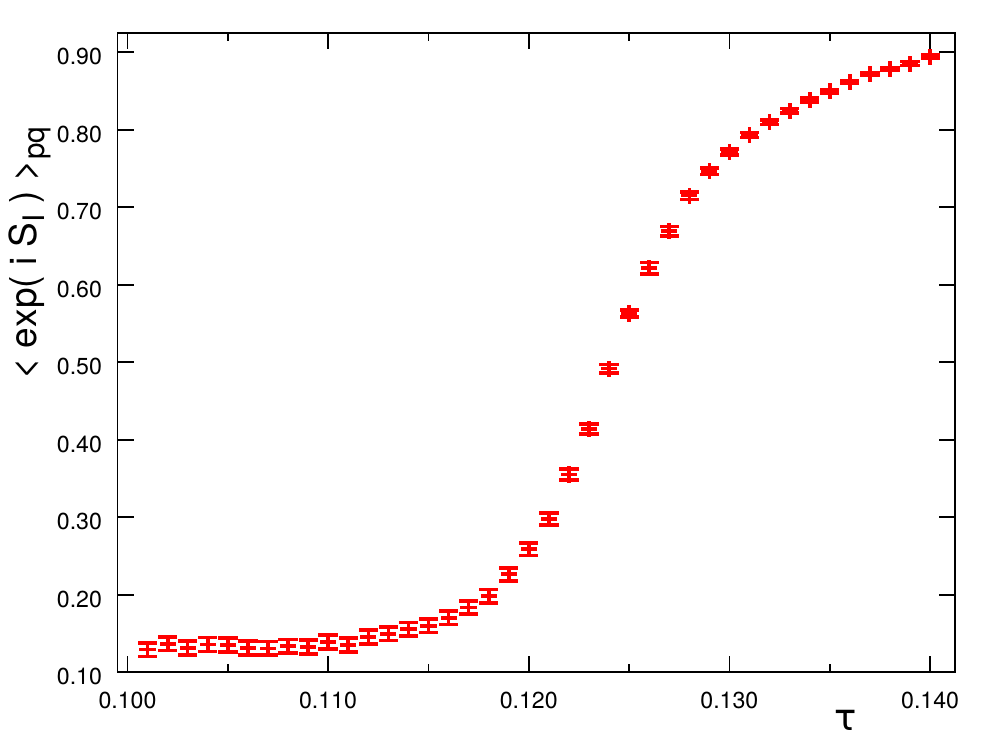}} 
\end{center}
\caption{Phase quenched expectation value of the complex phase. In the lhs.\ plot 
we show the results for $\kappa = 0.005$ and $\tau = 0.13$ as a function of $\mu$ 
for our $8^3$ and $12^3$ lattices, and on the rhs.\ the results for $\kappa = 0.005$, 
$\mu = 1.0$, and $8^4$ as a function of $\tau$.}
\label{fig:results_phase}
\end{figure}

We finalize our discussion of observables in Fig.~\ref{fig:results_phase}, where we show
$\langle e^{i S_I} \rangle_{pq}$, the phase quenched expectation value of the complex
phase, which is a measure for the severeness of the complex action problem. 
The lhs.\ plot shows $\langle e^{i S_I} \rangle_{pq}$ as a function of $\mu$ at $\kappa = 0.005, \tau = 0.13$ for 
$8^3$ and $12^3$, while the rhs.\ plot shows $\langle e^{i S_I} \rangle_{pq}$ as a function of $\tau$ at 
$\kappa = 0.005, \mu = 1.0$ for $8^3$. In the lhs.\ plot one nicely sees that the transitory behavior which we also 
observe in the corresponding observables shown in Figs.~5 and 6 is also reflected in the behavior of 
$\langle e^{i S_I} \rangle_{pq}$, with the complex action problem being more severe in the vicinity of the
transition and then again for large $\mu$. We also find that, as expected, the complex action problem becomes more
severe when the volume increases.  When considered as a function of $\tau$ (rhs.\ plot) we find that 
$\langle e^{i S_I} \rangle_{pq}$ shows a monotonous behavior, matching the monotonous behavior seen in the 
corresponding observables in Fig.~8.

\section{Truncating the density}

It is obvious from Figs.~2, \ref{fig:rho-tau} and 7 that the weighted density $\rho(x)$ drops over many orders of 
magnitude when increasing $x$. Although the amount of decrease depends on the parameters, it is an 
interesting question whether it is necessary to evaluate the density $\rho(x)$ for all values of $x \in [0,x_{max}]$.  
Maybe for some parameter values it is possible to truncate the range where the density is computed and the error 
for observables from truncation is negligible compared to errors from other sources, in particular from the discretization 
of $x$ and from finite statistics. The basis for experimenting with truncation are Eqs.~(\ref{eq:obs_n}) and 
(\ref{eq:susce_n}) which make explicit that regions of $x$ where $\rho(x)$ is very small will contribute
little to observables. In this section we explore the idea of truncation and show that the numerical cost can be 
reduced considerably without significant changes of the results. 

Let us begin with an estimate of the truncation effect for the partition sum $Z$. By $x^*$ we denote the value $x$ 
where we truncate the density, i.e., we set $\rho(x) = 0$ for $x > x^*$. The partition sum changes by an amount 
$\delta Z$ which is given by
\begin{equation}
\delta  Z \; = \; 2 \!\! \int\limits_{x^*}^{x_{max}} \!\! dx \, \rho(x) \cos( 2 k \sinh \! \mu \, x)  \; .
\end{equation}  
A (very conservative) bound for this integral is given
by $| \delta Z | \leq 2 \, x_{max}  \times \max_{\, x \geq x^*} \rho(x)$.  
Considering that for reasonable lattice sizes $x_{max} = V 3 \sqrt{3}/2$ is of order 
$\mathcal{O}(10^3)-\mathcal{O}(10^4)$ we conclude that when the density is smaller than, e.g., 
$e^{-50} \approx 10^{-22}$ for all $x > x^*$, then the contribution of the values $x > x^*$ is negligible. A density 
smaller than $e^{-50}$ corresponds to $\ln \rho(x) = - L(x) < -50$. This is the quantity we show in Figs.~2, \ref{fig:rho-tau} 
and 7 and it is obvious that for some parameter sets the value of $-50$ is reached already at relatively small $x$.

\begin{figure}[t]
\centering
{\includegraphics[width=0.6\textwidth]{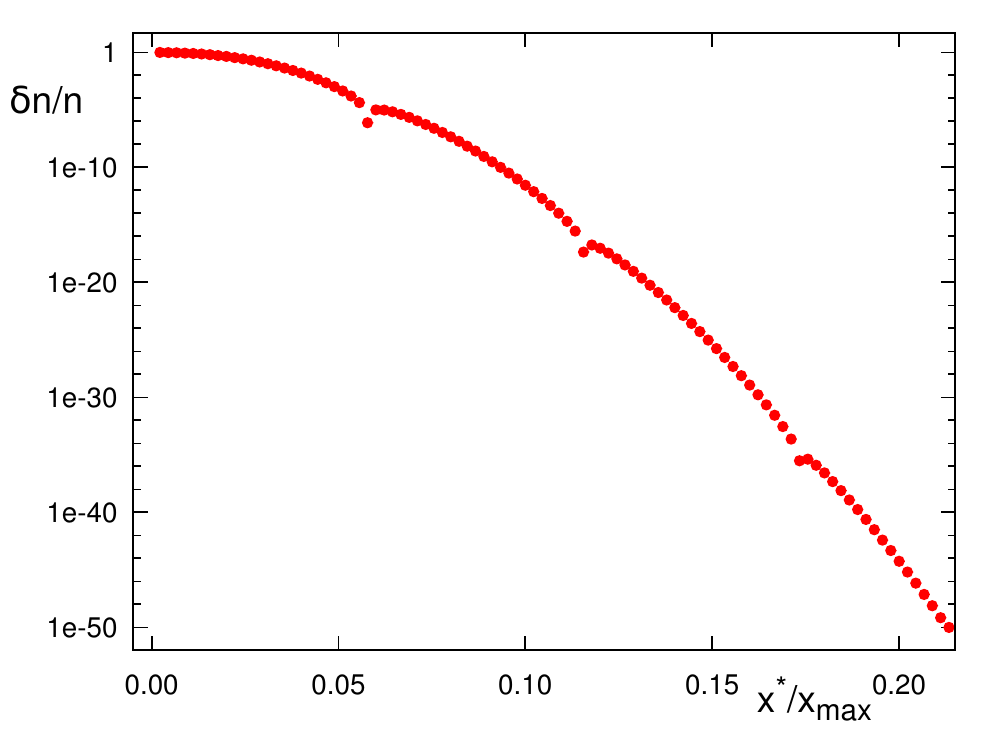}}
\caption{Example for the relative error $\delta n / n$ of the particle density $n$ due to truncation of 
the interval where the density $\rho(x)$ is evaluated. 
On the horizontal axis we show the position $x^*$ of the cut in units of $x_{max}$. The relative error drops
quickly already for relatively small values of $x^*/x_{max}$.}
\label{fig:cuttingDoS}
\end{figure}

After this first estimate for the partition sum we now study the effect of truncation for observables numerically.
More specifically, we define for the particle density $n$ the error
\begin{equation}
\delta n \;  = \;  \frac{1}{V} \, \frac{2}{Z-\delta Z} \, \int\limits_{x^*}^{x_{max}} \! \! dx \, \rho(x) \, \sin( 2 \kappa \sinh\! \mu \, x ) \, x  \; ,
\end{equation}
which is a simple generalization of $\delta Z$.
It is straightforward to give an upper bound for $\delta n$ in a similar way as for $\delta Z$. However, instead of 
discussing this bound, in Fig.~\ref{fig:cuttingDoS} we show the numerical results for the relative truncation error 
$\delta n / n$ as a function of $x^*/x_{max}$. The data are for $\mu=2.0$, $k=0.005$ and $\tau=0.066$, and we find 
that when cutting at $x^*/x_{max} = 0.2$, i.e., we take into account only 20 \% of the full range of $x$, the relative 
error has already dropped to $10^{-46}$. This plot impressively demonstrates that it is possible to considerably truncate 
the range of $x$ values where the density is evaluated and obtain an error which is completely negligible in comparison 
to other error sources. This saves a lot of numerical resources which can be invested in reducing these other
errors, e.g., by using smaller discretization intervals $\Delta_n$ or by increasing the statistics.

We stress again that the functional form of the density changes when varying the parameters (see Figs.~3 and 7)
and truncation can be applied only after checking whether the density is suitable. However, for that task 
we can again invoke 
preconditioning, i.e., the observation that a reasonable estimate for the overall behavior of the density can already be 
obtained from a first numerically inexpensive coarse estimate of $\rho(x)$ using few and large intervals $\Delta_n$.

\section{Summary and outlook}

Density of states techniques are an interesting approach to the simulation of lattice field theories that have a complex
action problem, e.g., theories with a chemical potential $\mu$. As discussed, the main challenge is to 
calculate the density of states $\rho(x)$ with sufficient accuracy. When evaluating observables the density is 
integrated over with a highly oscillating factor and the frequency of the oscillation increases exponentially with $\mu$. 
Thus, when one tries to reach reasonably large values of $\mu$, very high precision for $\rho(x)$ is mandatory.

In this paper we present the results of an exploratory implementation of the so-called Density of States Functional Fit 
Approach (DoS FFA) for the SU(3) spin model. The method uses a parameterization of the density $\rho(x)$ by 
an exponential of a piecewise linear function. The slopes of the linear pieces are computed with restricted vacuum
expectation values on the respective intervals. These restricted vacuum expectation values depend on a free parameter 
$\lambda$ and can be computed with standard Monte Carlo techniques. In each interval the functional form 
of the restricted vacuum expectation as function of $\lambda$ is known and depends only on the slope determining 
$\rho(x)$ on that interval. The slope then is obtained via a simple one-parameter fit to all Monte Carlo data and we 
show that the quality of the fit can be used as a self-consistency check of the method. 
From the slopes one determines $\rho(x)$ and with this density the observables.

We introduce a strategy which we refer to as ''preconditioning'': we show that already a coarse parameterization 
and low statistics are sufficient to get a good estimate for the overall behavior of the density. This information can then 
be used to optimize the interval size $\Delta_n$ 
for the piecewise linear parametrization of the exponent of $\rho(x)$, to determine
suitable ranges for the values of $\lambda$ to be used, and to assess a possible truncation of the density, a step which 
we illustrate to have a great potential for saving computer time that can be better used to reduce the error from 
discretization effects of $\rho(x)$ or from finite statistics.

We use the dual formulation of the SU(3) spin model for generating reference data for assessing the 
DoS FFA method, which is possible because the dual formulation is free of the complex action problem. 
In particular we study the particle number density and the corresponding susceptibility and find that the DoS FFA results 
match the dual simulation reference data for a surprisingly large range of $\mu$. 
Part of the reason for this success is the fact
that the shape of the density changes in a 
crossover transition that is crossed when increasing $\mu$, such that $\rho(x)$ drops faster with $x$,
a feature that is beneficial for the accuracy of the DoS FFA.
 
Encouraged by the success of the DoS FFA in the SU(3) spin model and the $\mathds{Z}_3$ model, studied in an earlier 
paper, we are currently working on developing the method further towards QCD. The essential steps will be the change 
from SU(3) spins to SU(3) gauge fields, and the formulation and implementation for theories with fermions. For both 
these issues we make progress and expect that the DoS FFA can be formulated for non-abelian gauge fields 
interacting with fermions as a generalization of the implementation presented here.

\vskip5mm
\noindent
{\bf Acknowledgements:} The authors thank Ydalia Delgado Mercado, Kurt Langfeld, Alexander Lehmann, 
and Biagio Lucini for many interesting discussions. This work is supported by the FWF DK W1203 
{\sl ''Hadrons in Vacuum, Nuclei and Stars''}, the FWF Grant.\ Nr.\ I 1452-N27, and partly also by 
DFG TR55, {\sl ''Hadron Properties from Lattice QCD''}.


\begin{thebibliography}{12}

\bibitem{Gocksch:1987nt}
  A.~Gocksch, P.~Rossi and U.M.~Heller,
  Phys.\ Lett.\ B {\bf 205} (1988) 334.

\bibitem{Gocksch:1988iz}
  A.~Gocksch,
  Phys.\ Rev.\ Lett.\  {\bf 61} (1988) 2054.

\bibitem{Schmidt:2005ap}
  C.~Schmidt, Z.~Fodor and S.D.~Katz,
  PoS LAT {\bf 2005} (2006) 163
  [hep-lat/0510087].

\bibitem{Fodor:2007vv}
  Z.~Fodor, S.D.~Katz and C.~Schmidt,
  JHEP {\bf 0703} (2007) 121   
  [hep-lat/0701022].

\bibitem{Ejiri:2007ga}
  S.~Ejiri,
  Phys.\ Rev.\ D {\bf 77} (2008) 014508
  [arXiv:0706.3549 [hep-lat]].

\bibitem{Ejiri:2012ng}
  S.~Ejiri {\it et al.} [WHOT-QCD Collaboration],
  Central Eur.\ J.\ Phys.\  {\bf 10} (2012) 1322
  [arXiv:1203.3793 [hep-lat]].

\bibitem{PhysRevLett.86.2050}
  F.~Wang and D.P.~Landau,
  Phys.\ Rev.\ Lett.\  {\bf 86} (2001)  2050
  [cond-mat/0011174 [cond-mat.stat-mech]].

\bibitem{Langfeld:2012ah}
  K.~Langfeld, B.~Lucini and A.~Rago,
  Phys.\ Rev.\ Lett.\  {\bf 109} (2012) 111601
  [arXiv:1204.3243 [hep-lat]].

\bibitem{Langfeld:2013xbf}
  K.~Langfeld and J.M.~Pawlowski,
  Phys.\ Rev.\ D {\bf 88} (2013) 071502
  [arXiv:1307.0455].
 
\bibitem{Langfeld:2014nta}
  K.~Langfeld and B.~Lucini,
  Phys.\ Rev.\ D {\bf 90} (2014) 094502
  [arXiv:1404.7187 [hep-lat]].

\bibitem{Langfeld:2015fua}
  K.~Langfeld, B.~Lucini, R.~Pellegrini and A.~Rago,
  Eur.\ Phys.\ J.\ C {\bf 76} (2016) 306
  [arXiv:1509.08391 [hep-lat]].

\bibitem{Garron:2016noc}
  N.~Garron and K.~Langfeld,
  arXiv:1605.02709 [hep-lat].

\bibitem{Gattringer:2016kco}
  C.~Gattringer and K.~Langfeld,
  arXiv:1603.09517 [hep-lat].

\bibitem{Mercado:2014dva}
  Y.~Delgado Mercado, P.~T\"orek and C.~Gattringer,
  PoS LATTICE {\bf 2014} (2015) 203
  [arXiv:1410.1645 [hep-lat]].
 
\bibitem{Gattringer:2015lra}
  C.~Gattringer and P.~T\"orek,
  Phys.\ Lett.\ B {\bf 747} (2015) 545
  [arXiv:1503.04947 [hep-lat]].

\bibitem{Gattringer:2015eey}
  C.~Gattringer, M.~Giuliani, A.~Lehmann and P.~T\"orek,
  POS LATTICE {\bf 2015} (2016) 194
  [arXiv:1511.07176 [hep-lat]].
 
\bibitem{Gattringer:2011gq}
  C.~Gattringer,
  Nucl.\ Phys.\ B {\bf 850} (2011) 242
  [arXiv:1104.2503 [hep-lat]].

\bibitem{Mercado:2012ue}
  Y.~Delgado Mercado and C.~Gattringer,
  Nucl.\ Phys.\ B {\bf 862} (2012) 737
  [arXiv:1204.6074 [hep-lat]].

\bibitem{Delgado:2012uh}
  Y.~Delgado and C.~Gattringer,
  Acta Phys.\ Polon.\ Supp.\  {\bf 5} (2012) 1033
  [arXiv:1208.1169 [hep-lat]].

 
\end{thebibliography}
\end{document}